\documentclass[dvips]{article}
\usepackage{supertabular,lscape,epsfig}
\usepackage{amssymb}
\usepackage{amsmath}

\DeclareSymbolFont{ppa}{OT1}{ppl}{m}{it}
\DeclareMathSymbol{\vv}{\mathalpha}{ppa}{'166}

\begin{document}

\newcommand{\dd}{\,{\rm d}}
\newcommand{\ie}{{\it i.e.},\,}
\newcommand{\etal}{{\it et al.\ }}
\newcommand{\eg}{{\it e.g.},\,}
\newcommand{\cf}{{\it cf.\ }}
\newcommand{\vs}{{\it vs.\ }}
\newcommand{\zdot}{\makebox[0pt][l]{.}}
\newcommand{\up}[1]{\ifmmode^{\rm #1}\else$^{\rm #1}$\fi}
\newcommand{\dn}[1]{\ifmmode_{\rm #1}\else$_{\rm #1}$\fi}
\newcommand{\upd}{\up{d}}
\newcommand{\uph}{\up{h}}
\newcommand{\upm}{\up{m}}
\newcommand{\ups}{\up{s}}
\newcommand{\arcd}{\ifmmode^{\circ}\else$^{\circ}$\fi}
\newcommand{\arcm}{\ifmmode{'}\else$'$\fi}
\newcommand{\arcs}{\ifmmode{''}\else$''$\fi}
\newcommand{\MS}{{\rm M}\ifmmode_{\odot}\else$_{\odot}$\fi}
\newcommand{\RS}{{\rm R}\ifmmode_{\odot}\else$_{\odot}$\fi}
\newcommand{\LS}{{\rm L}\ifmmode_{\odot}\else$_{\odot}$\fi}

\newcommand{\Abstract}[2]{{\footnotesize\begin{center}ABSTRACT\end{center}
\vspace{1mm}\par#1\par
\noindent
{~}{\it #2}}}

\newcommand{\TabCap}[2]{\begin{center}\parbox[t]{#1}{\begin{center}
  \small {\spaceskip 2pt plus 1pt minus 1pt T a b l e}
  \refstepcounter{table}\thetable \\[2mm]
  \footnotesize #2 \end{center}}\end{center}}

\newcommand{\TableSep}[2]{\begin{table}[p]\vspace{#1}
\TabCap{#2}\end{table}}

\newcommand{\FigCap}[1]{\footnotesize\par\noindent Fig.\  %
  \refstepcounter{figure}\thefigure. #1\par}

\newcommand{\TableFont}{\footnotesize}
\newcommand{\TableFontIt}{\ttit}
\newcommand{\SetTableFont}[1]{\renewcommand{\TableFont}{#1}}

\newcommand{\MakeTable}[4]{\begin{table}[htb]\TabCap{#2}{#3}
  \begin{center} \TableFont \begin{tabular}{#1} #4 
  \end{tabular}\end{center}\end{table}}

\newcommand{\MakeTableSep}[4]{\begin{table}[p]\TabCap{#2}{#3}
  \begin{center} \TableFont \begin{tabular}{#1} #4 
  \end{tabular}\end{center}\end{table}}

\newenvironment{references}%
{
\footnotesize \frenchspacing
\renewcommand{\thesection}{}
\renewcommand{\in}{{\rm in }}
\renewcommand{\AA}{Astron.\ Astrophys.}
\newcommand{\AAS}{Astron.~Astrophys.~Suppl.~Ser.}
\newcommand{\ApJ}{Astrophys.\ J.}
\newcommand{\ApJS}{Astrophys.\ J.~Suppl.~Ser.}
\newcommand{\ApJL}{Astrophys.\ J.~Letters}
\newcommand{\AJ}{Astron.\ J.}
\newcommand{\IBVS}{IBVS}
\newcommand{\PASP}{P.A.S.P.}
\newcommand{\Acta}{Acta Astron.}
\newcommand{\MNRAS}{MNRAS}
\renewcommand{\and}{{\rm and }}
\section{{\rm REFERENCES}}
\sloppy \hyphenpenalty10000
\begin{list}{}{\leftmargin1cm\listparindent-1cm
\itemindent\listparindent\parsep0pt\itemsep0pt}}%
{\end{list}\vspace{2mm}}

\def\TYLDA{~}
\newlength{\DW}
\settowidth{\DW}{0}
\newcommand{\dw}{\hspace{\DW}}

\newcommand{\refitem}[5]{\item[]{#1} #2%
\def\REFARG{#3}\ifx\REFARG\TYLDA\else, {\it#3}\fi
\def\REFARG{#4}\ifx\REFARG\TYLDA\else, {\bf#4}\fi
\def\REFARG{#5}\ifx\REFARG\TYLDA\else, {#5}\fi.}

\newcommand{\Section}[1]{\section{\hskip-6mm.\hskip3mm#1}}
\newcommand{\Subsection}[1]{\subsection{#1}}
\newcommand{\Acknow}[1]{\par\vspace{5mm}{\bf Acknowledgements.} #1}
\pagestyle{myheadings}

\newfont{\bb}{ptmbi8t at 12pt}
\newcommand{\xrule}{\rule{0pt}{2.5ex}}
\newcommand{\xxrule}{\rule[-1.8ex]{0pt}{4.5ex}}
\def\thefootnote{\fnsymbol{footnote}}
\begin{center}
{\Large\bf The Optical Gravitational Lensing Experiment.
\vskip1pt
Search for Planetary and Low-Luminosity 
\vskip1pt
Object Transits in the Galactic Disk. 
\vskip3pt
Results of 2001 Campaign\footnote{Based on observations obtained
with the 1.3~m Warsaw telescope at the Las Campanas Observatory of the
Carnegie Institution of Washington.}}
\vskip.6cm
{\bf A.~~U~d~a~l~s~k~i$^1$, ~~B.~~P~a~c~z~y~{\'n}~s~k~i$^2$,
~~K.~~\.Z~e~b~r~u~\'n$^1$, ~~M.~~S~z~y~m~a~{\'n}~s~k~i$^1$,
~~M.~~K~u~b~i~a~k$^1$, ~~I.~~S~o~s~z~y~\'n~s~k~i$^1$,
~~O.~~S~z~e~w~c~z~y~k$^1$, ~~\L.~~W~y~r~z~y~k~o~w~s~k~i$^1$,
~~and~~G.~~P~i~e~t~r~z~y~\'n~s~k~i$^{3,1}$}
\vskip2mm
$^1$Warsaw University Observatory, Al.~Ujazdowskie~4, 00-478~Warszawa, Poland\\
e-mail: (udalski,zebrun,msz,mk,soszynsk,szewczyk,wyrzykow,pietrzyn)@astrouw.edu.pl\\
$^2$ Princeton University Observatory, Princeton, NJ 08544-1001, USA\\
e-mail: bp@astro.princeton.edu\\
$^3$ Universidad de Concepci{\'o}n, Departamento de Fisica,
Casilla 160--C, Concepci{\'o}n, Chile
\end{center}

\vspace*{7pt} 
\Abstract{We present results of an extensive photometric search for
planetary and low-luminosity object transits in the Galactic disk stars
commencing the third phase of the Optical Gravitational Lensing
Experiment -- OGLE-III. Photometric observations of three fields in the
direction of the Galactic center (800 epochs per field) were collected 
on 32 nights during time interval of 45 days. Out of the total of 5
million stars monitored, about 52~000 Galactic disk stars with photometry
better than 1.5\%  were analyzed for flat-bottomed eclipses with the
depth smaller than 0.08~mag.

Altogether 46 stars with transiting low-luminosity objects were
detected. For 42 of them multiple transits were observed, a total of
185, allowing orbital period determination. Transits in two objects:
OGLE-TR-40 and OGLE-TR-10, with the radii ratio of about 0.14 and
estimate of the radius of the companion $1.0{-}1.5~{\rm R}_{\rm Jup}$,
resemble the well known planetary transit in HD~209458.  

The sample was selected by the presence of apparent transits only, with no 
knowledge on any other properties. Hence, it is very well suited for general 
study of low-luminosity objects. The transiting objects may be Jupiters, brown 
dwarfs, or M dwarfs. Future determination of the amplitude of radial velocity 
changes will establish their masses, and will confirm or refute the reality of 
the so called ``brown dwarf desert''. The low-mass stellar companions will 
provide new data needed for the poorly known mass-radius relation for the 
lower main sequence. 

All photometric data are available to the astronomical community from
the OGLE Internet archive.}{} 

\Section{Introduction}
In the past decade astronomers witnessed a breathtaking progress in the
search for extrasolar planets. The  discovery of mysterious planet-like
objects around pulsar PSR~B1257+12 (Wolszczan and Frail 1992) was
followed by the detection of the first solar type planets around nearby
stars (Mayor and Queloz 1995, Marcy and Butler 1996). While the former
type system seems to be unique and its origin and status remain
mysterious, the solar type planets seem to be quite common in the solar
neighborhood: up to now about 80 planetary systems have been reported
(see {\it http://www.obspm.fr/planets} or {\it http://exoplanets.org} 
for the most recent information).  

Apart from PSR~B1257+12, whose discovery was based on radio wavelength
pulse time delay measurements, ultra-precise measurements of the radial
velocity shifts in the optical spectra of target stars were the main
successful method of planet detection. Such measurements provided
orbital periods and minimum masses of the companion (\cf Mazeh and
Zucker 2002, and references therein).

Extensive spectroscopic searches of nearby dwarf stars revealed a large
variety of planetary orbits.  A distinct class is a group of several
stars with Jupiter-mass planets orbiting their stars in several days
($a<0.1$~a.u.), and  called ``51~Peg-like'' planets or ``hot Jupiters''.
A very important feature of this class is a relatively large
probability that a planetary orbit has an inclination allowing a transit
in front of the star. While for our solar system such a probability
would be very small (Sackett 1999), it is about 10\% for ``51~Peg-like''
planets. The expected depth of a transit is small, $ \approx 1\% $ even
for ``Jupiters''. However, this is within reach of a careful
ground-based photometry. 

It is not surprising that the ``51~Peg-like'' planets were monitored
photometrically. At the end of 1999 a spectacular discovery of the first
transit of a planet around HD~209458 was announced by Henry \etal (2000)
and Charbonneau \etal (2000). Combined spectroscopic data and photometry
of the transit allow to determine the most important planet parameters,
its mass and radius: $0.7~{\rm M}_{\rm Jup}$ and $1.4~{\rm R}_{\rm
Jup}$, respectively, for HD~209458  (Brown \etal 2001).

The discovery of HD~209458 transit proved that photometry may be used in
search for planets. However, due to a small probability of transits, a
photometric survey must monitor a large number of stars with  high
photometric accuracy every $\approx 10$~minutes for many nights.  The
most promising areas of the sky are the dense stellar regions like the
Galactic disk, globular clusters etc.  Ground-based accurate photometry
used to be difficult in crowded fields. It has recently become possible
thanks to the technique called ``image subtraction'' or ``difference
image analysis'' (DIA) developed by Alard and Lupton (1998) and Alard
(2000).

Several planetary transit monitoring programs are presently underway.
Gilli\-land \etal (2000) monitored 47~Tuc globular cluster with the HST
telescope. No transit was found in 34~000 stars monitored over the
period of 8.3~days. The same cluster was also monitored from the ground
(Sackett 2000, private communication). Other programs are focused on
open clusters fields (Quirrenbach \etal 2000, Street \etal 2000, PISCES
project -- Mochejska \etal 2002), Galactic disk fields (EXPLORE project
-- Mallen-Ornelas \etal 2002), or brighter stars -- STARE project (Brown
and Charbonneau 2000), VULCAN project (Borucki \etal 2001). Also space
missions for continuous photometric monitoring of nearby stars, capable
to detect Earth-like planet transits, like COROT (Deleuil \etal 2000)
and KEPLER ({\it http://www.kepler.arc.nasa.gov}) are planned. 
However, HD~209458 remains the only planetary transit system known.

The photometry alone cannot unambiguously distinguish between Jupiter
size planets and other low-luminosity objects: brown dwarfs and late
type M dwarfs, as all of them have the radii of the order of
0.1--0.2~\RS\ (1--2~R$_{\rm Jup}$). Thus,  a spectroscopic follow-up and
a measurement of the radial velocity amplitude of the stars is needed to
determine the masses of transiting companions.

Although detection of planets {\it via} transits is certainly the most
exciting possibility, detection of any other type of low-luminosity
objects is also very important.  There appears to exist so called
``brown dwarf desert'': absence of brown dwarfs with short period orbits
around stars (\cf Tabachnik and Tremaine 2001, and references therein).
The mass-radius relation for the lower main sequence is poorly known,
and a serious discrepancy between the observations and models can be
seen (Torres and Ribas 2001,  O'Brien, Bond and Sion 2001, \cf Spruit
1982, Spruit and Weiss 1986, for a possible explanation). Summarizing,
photometric detection of transits, combined with the spectroscopic
follow-up, will provide valuable information about all  three types of
low-luminosity stellar companions.

The Optical Gravitational Lensing Experiment (OGLE) is a long term
observational program started almost a decade ago (Udalski \etal 1992,
Udalski, Kubiak and Szyma{\'n}ski 1997) with the initial goal of
detection of microlensing events {\it via} photometric monitoring of
millions of stars in the densest stellar regions of the sky. The OGLE
survey turned out to be very successful in discovering not only hundreds
of microlensing events (Udalski \etal 1993, Wo{\'z}niak \etal 2001) but
also providing huge amount of data on the variable (Udalski \etal 1999,
{\.Z}ebru{\'n} \etal 2001, Wo{\'z}niak \etal 2002) and non-variable
(Udalski \etal 1998, Udalski \etal 2000)  stars in the Galactic bulge
and Magellanic Clouds and contributing to many other fields. 

At the beginning of 2001, the OGLE project underwent its second major
hardware upgrade. A single chip {${2048\times2048}$} pixel CCD camera was
replaced by the second generation eight chip {$8192\times 8192$} pixel
CCD mosaic. Also the old data reduction pipeline was replaced with the
new one based on image subtraction. The data flow of the survey
increased by almost an order of magnitude, and  the accuracy of
photometry reached the millimagnitude level for the brightest stars.

The third phase of the OGLE survey, OGLE-III, started on June 12, 2001.
For the first 45 days three fields located in the densest stellar
regions in the direction of the Galactic center were monitored
continuously up to 35 times per night, on a total of 32 nights, for
detection of short time-scale stellar variability. These data are ideal
to search for planetary transits; this was one of the goals of the
campaign. Such a pilot search was undertaken to answer the question
whether planetary transit detection is possible and the program is worth
continuing in the next observing seasons. 

In this paper we present results of the OGLE-III search for planetary
and low-luminosity object transits in the  photometric data collected
during the 2001 campaign. We discovered 185 individual transit cases in
46 objects from about 52~000 Galactic disk stars for which the
photometric accuracy was 1.5\% or better.  These transits may be due to
``Jupiters'', brown dwarfs or M dwarfs.  The distinction will be
possible when a spectroscopic follow-up will be done.  The photometry of
all OGLE-III stars  with transits is available to the astronomical
community from the OGLE  Internet archive.

\vskip-9pt
\Section{Observational Data}
\vskip-9pt
Observations presented in this paper started the third phase of the OGLE
experiment, OGLE-III. They were collected with the 1.3-m Warsaw
telescope at the Las Campanas Observatory, Chile, (operated by the
Carnegie Institution of Washington) equipped with the new large field
CCD mosaic camera, consisting of eight ${2048\times4096}$ pixel SITe
ST002A detectors. More detailed description of the OGLE-III instrumental
system will be presented in the forthcoming paper. Here, we only
highlight the most important features of the OGLE-III second generation
camera and data reduction system. 

The pixel size of each of the detectors is 15~$\mu$m giving the 0.26 
arcsec/pixel scale at the focus of the Warsaw telescope. The full field
of view of the camera is about $35\times 35.5$ arcmins. The reading of
the entire array takes about 98 seconds and the size of a single full
frame is about 137~MB.  The gain of each chip is adjusted to be about
1.3~e$^-$/ADU with the readout noise from about 6 to 9~e$^-$, depending
on the chip. The well depth of chips reaches from 60000~e$^-$ to
80000~e$^-$. Contrary to the OGLE-II instrumentation, where the vast
majority of images were taken in the ``driftscan'' mode, the new camera
only works in the classical ``still-frame'' mode.

The vast majority of photometric data were taken during 32 nights
spanning 45 days from June~12, 2001. After that period single
observations of presented fields were done once every few nights up to
the end of 2001 Galactic bulge season in October~2001. Equatorial
coordinates (J2000.0) of the observed fields are listed in Table~1. Each
of the fields overlaps slightly with the neighboring one for calibration
purposes.
\MakeTable{lcc}{12.5cm}{Equatorial coordinates of the planetary transit fields}
{
\hline
\noalign{\vskip3pt}
\multicolumn{1}{c}{Field} & RA (J2000) & DEC (J2000)\\
\hline
\noalign{\vskip3pt}
BLG100  &  17\uph51\upm00\ups & $-29\arcd59\arcm45\arcs$ \\
BLG101  &  17\uph53\upm40\ups & $-29\arcd49\arcm50\arcs$ \\
BLG102  &  17\uph56\upm20\ups & $-29\arcd30\arcm50\arcs$ \\
\hline}

Almost all observations were made in the {\it I}-band filter. The
exposure time of each image was set to 120 seconds. Altogether about 800
epochs were collected for each field during the 2001 season.
Additionally, several {\it V}-band frames were also made to have color
information for the observed objects. {\it V}-band images were exposed
for 150~seconds.

Because of extremely high stellar density of the observed fields the
observations were carried out only during good seeing conditions and
were suspended when the seeing exceeded 1.8~arcsec. The median seeing of
the presented dataset is 1.2~arcsec.

\Section{Data Reduction}
Similarly to the previous phases of the OGLE survey, data reduction in
OGLE-III is supposed to be performed at the telescope in almost real
time. However, during the 2001 campaign only the first part of the data
reduction pipeline, namely the initial reduction like de-biasing and
flat-fielding, was implemented. Photometric reductions of collected
images were performed off-line, however, with the same pipeline which
will be implemented on-line at the OGLE telescope in 2002 observing
season.

The initial reduction procedures were based on the IRAF\footnote{IRAF is
distributed by National Optical Observatories, which is operated by the
Association of Universities for Research in Astronomy, Inc., under
cooperative agreement with National Science Foundation.} {\sc CCDRED}
package routines and were performed automatically when the new frame
arrived from the telescope. To make data handling easier, each chip of
the mosaic was treated separately. Flat-fielded data, were compressed with
the RICE algorithm and stored on a few hundred GB partition of the hard
disk, and finally dumped to the HP Ultrium tape (about 200~GB of raw
data per tape).

The photometric data reduction was based on the difference image
analysis (DIA) method (Alard and Lupton 1998, Alard 2000). The core of
the system -- image subtraction procedures were based on the
implementation of this method by Wo{\'z}niak (2000), with many software
modifications for better performance and stability. Contrary to the
Wo{\'z}niak (2000) approach (which in principle works off-line, well
after all images of the field are collected, while presented data
pipeline is designed to analyze photometry on-the-fly) photometry on the
subtracted image is performed for all objects identified earlier in the
reference image at the position of their centroids. Subtracted image is
also analyzed for stars which apparently changed their brightness above
the preassigned threshold, so that the on-line information on the
current variability is obtained. Objects which cannot be
cross-identified with reference image stars on the subtracted image  are
treated as new objects in the current frame (although this list may
contain artifacts like small traces of non-perfectly removed cosmic-rays
etc.). The threshold for detection of the current variability is by
definition too high for triggering by very small amplitude variations like
transits. However, full photometric information on the brightness of all
objects in the image is preserved in the file with the photometry of all
objects.

Next, these files are used to feed the final photometric databases. We
use the same database software as during the previous stages of the OGLE
project (Szyma{\'n}ski and Udalski 1993) with minor modifications due to
somewhat different data input format. Separate databases are created for
each chip in a given field, thus the complete set for a field consists
of eight databases of about 1~GB each in the case of the Galactic bulge
data collected in 2001 season.

The main advantage of the DIA photometry over the classical PSF approach
is a much higher accuracy. This is achieved due to two main reasons.
First, variability is measured very precisely on an almost empty
subtracted image. Secondly, a much better quality profile photometry for
constant part of the flux can be obtained from the reference image which
has  much higher S/N than a single frame, as it is obtained by stacking
several (more than 10 in our case) single images.   A large number of
stars in the observed fields assures a very good match between the
reduced and reference images before the subtraction is done. 

Wo{\'z}niak (2000) noted a significantly better quality of photometry
obtained with the DIA over the standard OGLE-II data reduction pipeline
based on profile photometry. The comparison of the DIA results for OGLE-III
frames indicates that the quality of OGLE-III data is even superior over
the DIA OGLE-II photometry. This is certainly due to a better sampling
(0.26 \vs 0.42 arcsec/pixel) and better quality of images obtained in the
``still-frame'' mode. For the brightest stars {\it rms} of photometric
measurements from all, more than 800, epochs is at a few millimagnitudes
level.

Observations in the 2001 season were focused on a detection of rapid
variability, and no calibration images with standard stars were
collected. Therefore it was not possible to tie directly the photometry
to the standard system with a very high precision. However, to have a
crude estimate of the zero point of our photometry we compared our
instrumental magnitude scale with the well calibrated photometry of
OGLE-II fields (Udalski \etal 2002, in preparation). In many cases
OGLE-II and OGLE-III fields overlap, hence a comparison was possible. If
the object was observed during the OGLE-II phase we shifted OGLE-III
magnitudes to fit OGLE-II calibrated values. In this case the zero point
accuracy is better  than 0.05~mag. If there was no OGLE-II/OGLE-III
overlap then an average shift resulting from a comparison of other stars
in the same field was applied.  We believe that the zero point accuracy
of our magnitude scale is  about 0.1~mag in this case. It should be
noted that although OGLE-II photometric measurements exist for large
number of our transit objects, they are not suitable for search or even
confirmation of transits because of much smaller number of observations,
non-appropriate sampling (one observation per 1--3 days) and much worse
photometric quality. 

Astrometric solution for the observed fields was performed in identical
manner as in the OGLE-II photometric maps case (Udalski \etal 1998), \ie
by cross-identification of about 2000 brightest stars in each chip
reference image with the Digitized Sky Survey images of the same part of
the sky. Then the transformation between OGLE-III pixel grid and
equatorial coordinates of the DSS (GSC) astrometric system was
calculated. The systematic error of the latter can be up to about
0.7~arcsec, while the internal error of the transformation is about
0.2~arcsec.

\begin{figure}[htb]
\vglue2mm
\includegraphics[width=10.8cm, bb=30 45 410 500]{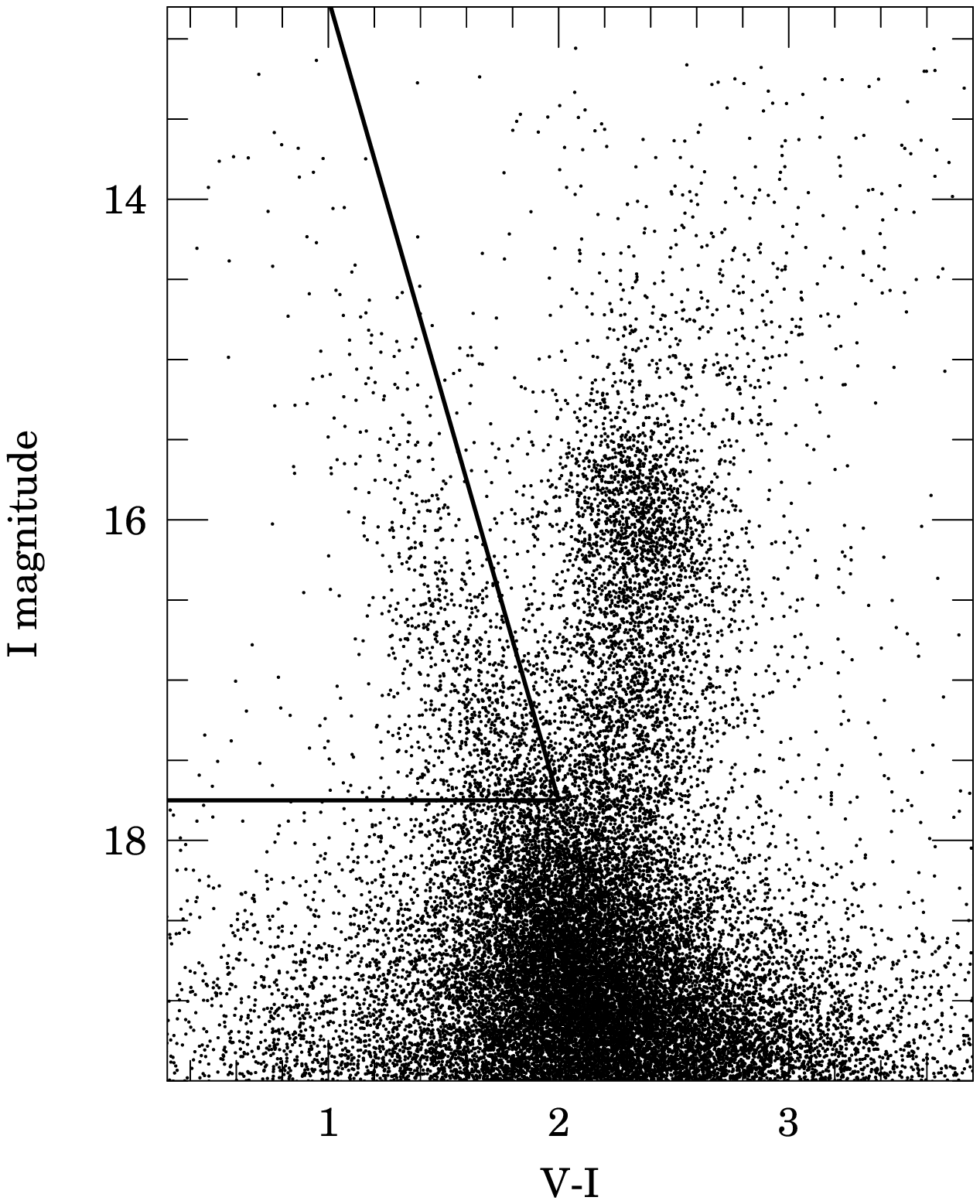}
\FigCap{Color-Magnitude diagram of the BLG102.4 field. Only 20\% of stars
are plotted for clarity. Solid line limits the Galactic disk stars
searched for transits. Accuracy of the zero points of color and {\it
I}-band magnitude scale is about 0.1~mag.}
\end{figure}

\Section{Selection of Transit Candidates}
A huge number of the observed stars requires a fully automatic method of
selection of transit candidates for the work to be effective. We
performed  our search for transits as follows. The three observed fields
contain more than 5 million stars, located both in the Galactic disk and
in the Galactic bulge. Therefore, we first limited the sample to the
stars located in the disk. To do this we observed our fields several
times in the {\it V}-band. A preliminary reduction of these images was
used to construct the photometric databases, similar to the main {\it
I}-band databases.  Next, we prepared the color-magnitude diagrams
(CMDs) of the observed fields. One of such diagrams is presented in
Fig.~1.  

For brighter stars the separation of the disk main sequence stars
(forming a clear sequence in Fig.~1) from the Galactic bulge stars (red
giant and subgiant stars) is clear.  Hence, further analysis was
restricted to the stars from the region limited by bold lines in Fig.~1.

We further limited our sample to the stars with the standard deviation
of all observed magnitudes ${\leq 0.015}$~mag. To avoid contamination of
the standard deviation by outlying observations we ran $5\sigma$
clipping filter on the data prior to the calculations of the standard
deviation. We were left with about 52~000 stars in the three fields
which passed our initial filtering.

In the next step all these objects were subjected to the transit finding
procedure. We constructed an artificial error-less transit light curve
with an amplitude of 0.015~mag and the total duration of 0.03 phase.
Next, the observations of the analyzed star were folded with a trial
period and cross-correlated with the artificial light curve. If the
cross-correlation coefficient was larger than a preselected threshold
the star was marked as a candidate. The procedure was repeated for trial
periods in the range of 1--10~days with the step of ${0.0001\cdot P}$.

After the initial tune-up of the cross-correlation coefficient threshold
this algorithm worked very effectively. Practically all triggered cases
turned out to be small amplitude eclipsing objects and the contamination
by artifacts was minimal. The algorithm can easily be used to calibrate
the efficiency of our detection procedure by running it on artificially
prepared eclipsing light curves.  This, however,  has not been done yet. 
The algorithm is very sensitive even to single transit events during our
observing run, as there is practically always a  period (false, of
course) which mimics a regular eclipsing light curve with some gaps in
the observations. Such a light curve is easily detected by our search
algorithm.

The final selection of candidates was done by a careful visual inspection
of the photometric data, both in the day scale and phased. We left on
the main list of candidates only those stars which have a significant
probability of being true transits. We do not present here a large
number of small amplitude events caused by grazing eclipses (V-shape
eclipses, often with some shift of the secondary eclipse phase) or
somewhat deeper (${>0.1}$~mag) cases which occasionally passed our
filters. However, one has to remember that a regular faint binary with a
flat-bottomed eclipse may blend with a much brighter star, to mimic
low amplitude transit. In our very dense stellar fields the probability
of blending is non-negligible.  Future spectroscopy will resolve this
possible problem.

The final periods of our candidates were found by a careful examination
of the eclipse light curve, by minimizing dispersion in the eclipse
phases. For some objects only uncertain information on periodicity could
be found, because only two or so, often incomplete eclipses were
registered. When a small number of transits was registered we always
adopted the shortest period consistent with our remaining data. The
formal accuracy of the period is of the order of ${2\cdot10^{-4} P}$. In
a few cases we observed only one transit event so that nothing can be
said at this stage about the periodicity of these object.

\Section{Results of 2001 Campaign}
Forty six transit stars were found in the 2001 campaign data among 52~000
searched Galactic disk objects. The star entered the final list of
transit objects when the amplitude of transit was smaller than 0.08~mag.
With a completely dark transiting object such an amplitude still
corresponds to $1.4~{\rm R}_{\rm Jup}$ if the stellar radius is half the
solar.

We present the results in the form of a catalog and atlas of light
curves. Table~2 contains basic data for each star: abbreviation in the
form OGLE-TR-NN, equatorial coordinates (J2000.0 epoch), orbital period,
epoch of mid-eclipse, {\it I}-band magnitude and an approximate ${V-I}$
color outside eclipse, the depth of eclipse, number of transits
observed ($N_{\rm tr}$),  and remarks. OII acronym in the remarks column
indicates that the object was observed in the OGLE-II phase.  For these
stars both -- the mean {\it I}-band magnitude and ${V-I}$ color from
OGLE-II photometric maps were used to set the zero point of photometry,
and both values are accurate to better than 0.05~mag. In the remaining
cases the accuracy of the magnitude scale and $V-I$ color is not better
than 0.1~mag.   

\renewcommand{\arraystretch}{1.11}
\renewcommand{\TableFont}{\scriptsize} 
\MakeTableSep{l@{\hspace{6pt}}
c@{\hspace{5pt}}c@{\hspace{6pt}}c@{\hspace{5pt}}c@{\hspace{4pt}}
c@{\hspace{4pt}}c@{\hspace{4pt}}c@{\hspace{4pt}}c@{\hspace{4pt}}l}{12.5cm}
{OGLE-III planetary and low-luminosity object transits} 
{\hline
\noalign{\vskip4pt} Name       & RA (J2000)  & DEC (J2000) &   $P$    &
$T_0$         & $I$    &$V-I$ &$\Delta I$  &  $N_{\rm tr}$ &Rem.\\
           &             &             & [days]   & --2452000     & [mag]  &[mag] &[mag]  &\\ 
\noalign{\vskip4pt}
\hline
\noalign{\vskip4pt}
OGLE-TR-1  & $17\uph51\upm10\zdot\ups02$ & $-30\arcd16\arcm46\zdot\arcs2$ &   1.6009 & 61.14367 & 15.657 & 1.30 & 0.043 & 4 &  \\
OGLE-TR-2  & $17\uph51\upm24\zdot\ups24$ & $-30\arcd14\arcm05\zdot\arcs9$ &   2.8131 & 61.86463 & 14.177 & 0.85 & 0.019 & 5 &  \\
OGLE-TR-3  & $17\uph51\upm48\zdot\ups95$ & $-30\arcd13\arcm25\zdot\arcs1$ &   1.1899 & 60.22765 & 15.567 & 1.00 & 0.019 &10 &  \\ 
OGLE-TR-4  & $17\uph51\upm14\zdot\ups55$ & $-30\arcd03\arcm28\zdot\arcs3$ &   2.6188 & 60.21812 & 14.715 & 1.40 & 0.065 & 5 &  \\ 
OGLE-TR-5  & $17\uph51\upm49\zdot\ups39$ & $-30\arcd01\arcm44\zdot\arcs3$ &   0.8082 & 60.47118 & 14.883 & 0.90 & 0.043 & 8 &  \\ 
OGLE-TR-6  & $17\uph51\upm03\zdot\ups07$ & $-29\arcd55\arcm49\zdot\arcs8$ &   4.5487 & 61.05651 & 15.356 & 1.70 & 0.053 & 4 &  \\ 
OGLE-TR-7  & $17\uph52\upm08\zdot\ups63$ & $-29\arcd56\arcm12\zdot\arcs7$ &   2.7179 & 61.42566 & 14.800 & 1.10 & 0.034 & 5 & OII \\ 
OGLE-TR-8  & $17\uph52\upm18\zdot\ups57$ & $-29\arcd56\arcm24\zdot\arcs6$ &   2.7152 & 61.01604 & 15.648 & 1.20 & 0.048 & 4 & OII \\ 
OGLE-TR-9  & $17\uph51\upm50\zdot\ups88$ & $-29\arcd54\arcm43\zdot\arcs4$ &   3.2687 & 61.42444 & 14.012 & 0.85 & 0.048 & 4 &  \\ 
OGLE-TR-10 & $17\uph51\upm28\zdot\ups25$ & $-29\arcd52\arcm34\zdot\arcs9$ &   3.1037 & 60.88835 & 14.930 & 0.85 & 0.022 & 4 &  \\ 
OGLE-TR-11 & $17\uph50\upm04\zdot\ups96$ & $-29\arcd57\arcm40\zdot\arcs7$ &   1.6154 & 60.37108 & 16.006 & 1.99 & 0.053 & 7 & OII \\ 
OGLE-TR-12 & $17\uph50\upm49\zdot\ups55$ & $-30\arcd01\arcm05\zdot\arcs6$ &   5.7721 & 61.53515 & 14.671 & 1.22 & 0.038 & 2 & OII \\ 
OGLE-TR-13 & $17\uph50\upm55\zdot\ups41$ & $-30\arcd14\arcm51\zdot\arcs2$ &   5.8534 & 64.68849 & 13.892 & 0.85 & 0.030 & 2 &  \\ 
OGLE-TR-14 & $17\uph54\upm33\zdot\ups92$ & $-30\arcd01\arcm31\zdot\arcs5$ &   7.7978 & 62.28715 & 13.064 & 0.81 & 0.034 & 3 & OII \\ 
OGLE-TR-15 & $17\uph54\upm52\zdot\ups30$ & $-29\arcd58\arcm20\zdot\arcs4$ &   4.8746 & 61.18221 & 13.226 & 0.93 & 0.026 & 4 & OII \\ 
OGLE-TR-16 & $17\uph54\upm08\zdot\ups96$ & $-29\arcd47\arcm39\zdot\arcs3$ &   2.1386 & 60.33385 & 13.517 & 0.99 & 0.026 & 4 & OII \\ 
OGLE-TR-17 & $17\uph54\upm23\zdot\ups53$ & $-29\arcd45\arcm58\zdot\arcs4$ &   2.3171 & 62.35748 & 16.218 & 1.26 & 0.034 & 4 & OII \\ 
OGLE-TR-18 & $17\uph54\upm16\zdot\ups46$ & $-29\arcd43\arcm11\zdot\arcs9$ &   2.2280 & 61.07501 & 16.010 & 1.25 & 0.043 & 6 & OII \\ 
OGLE-TR-19 & $17\uph54\upm33\zdot\ups38$ & $-29\arcd44\arcm37\zdot\arcs8$ &   5.2821 & 61.89798 & 16.351 & 1.57 & 0.065 & 2 & OII \\ 
OGLE-TR-20 & $17\uph53\upm51\zdot\ups72$ & $-29\arcd41\arcm53\zdot\arcs8$ &   4.2835 & 63.87664 & 15.405 & 1.24 & 0.059 & 3 & OII \\ 
OGLE-TR-21 & $17\uph54\upm47\zdot\ups04$ & $-29\arcd41\arcm17\zdot\arcs4$ &   6.8925 & 66.11971 & 15.585 & 1.51 & 0.043 & 3 & OII \\ 
OGLE-TR-22 & $17\uph54\upm38\zdot\ups63$ & $-29\arcd38\arcm32\zdot\arcs0$ &   4.2750 & 61.48734 & 14.546 & 1.14 & 0.084 & 3 & OII \\ 
OGLE-TR-23 & $17\uph54\upm35\zdot\ups05$ & $-29\arcd38\arcm50\zdot\arcs7$ &   3.2866 & 61.79082 & 16.394 & 1.21 & 0.059 & 2 & OII \\ 
OGLE-TR-24 & $17\uph53\upm04\zdot\ups49$ & $-29\arcd38\arcm30\zdot\arcs2$ &   5.2821 & 63.80639 & 14.847 & 1.20 & 0.053 & 2 & OII \\ 
OGLE-TR-25 & $17\uph52\upm45\zdot\ups41$ & $-29\arcd35\arcm12\zdot\arcs1$ &   2.2181 & 62.13473 & 15.274 & 1.26 & 0.038 & 5 & OII \\ 
OGLE-TR-26 & $17\uph53\upm21\zdot\ups22$ & $-29\arcd35\arcm38\zdot\arcs8$ &   2.5389 & 61.49645 & 14.784 & 1.24 & 0.053 & 4 & OII \\ 
OGLE-TR-27 & $17\uph53\upm36\zdot\ups77$ & $-29\arcd34\arcm29\zdot\arcs5$ &   1.7149 & 61.13038 & 15.716 & 1.37 & 0.026 & 6 & OII \\ 
OGLE-TR-28 & $17\uph52\upm46\zdot\ups39$ & $-29\arcd45\arcm13\zdot\arcs9$ &   3.4051 & 63.10716 & 16.433 & 1.44 & 0.053 & 3 & OII \\ 
OGLE-TR-29 & $17\uph52\upm18\zdot\ups53$ & $-29\arcd56\arcm24\zdot\arcs7$ &   2.7159 & 61.01619 & 15.648 & 1.20 & 0.048 & 4 & OII \\ 
OGLE-TR-30 & $17\uph52\upm48\zdot\ups58$ & $-30\arcd00\arcm30\zdot\arcs1$ &   2.3650 & 61.92604 & 14.926 & 1.09 & 0.038 & 6 & OII \\ 
OGLE-TR-31 & $17\uph53\upm22\zdot\ups69$ & $-29\arcd59\arcm23\zdot\arcs1$ &   1.8832 & 60.79767 & 14.335 & 1.06 & 0.030 & 7 & OII \\ 
OGLE-TR-32 & $17\uph56\upm47\zdot\ups53$ & $-29\arcd42\arcm42\zdot\arcs0$ &   1.3433 & 60.35064 & 14.853 & 0.95 & 0.034 & 7 &  \\
OGLE-TR-33 & $17\uph56\upm41\zdot\ups19$ & $-29\arcd40\arcm05\zdot\arcs3$ &   1.9533 & 60.54289 & 13.711 & 0.95 & 0.034 & 2 &  \\
OGLE-TR-34 & $17\uph56\upm44\zdot\ups90$ & $-29\arcd40\arcm34\zdot\arcs5$ &   8.5810 & 62.74970 & 15.995 & 1.45 & 0.048 & 3 &  \\ 
OGLE-TR-35 & $17\uph57\upm16\zdot\ups01$ & $-29\arcd35\arcm30\zdot\arcs5$ &   1.2599 & 60.98942 & 13.267 & 0.80 & 0.030 & 7 &  \\ 
OGLE-TR-36 & $17\uph57\upm38\zdot\ups02$ & $-29\arcd35\arcm17\zdot\arcs2$ &   6.2516 & 62.37694 & 15.767 & 1.45 & 0.059 & 2 &  \\ 
OGLE-TR-37 & $17\uph57\upm30\zdot\ups11$ & $-29\arcd28\arcm43\zdot\arcs6$ &   5.7197 & 60.57910 & 15.184 & 1.40 & 0.030 & 2 &  \\ 
OGLE-TR-38 & $17\uph56\upm21\zdot\ups17$ & $-29\arcd24\arcm00\zdot\arcs4$ &   4.1015 & 62.50097 & 14.674 & 0.60 & 0.048 & 3 &  \\ 
OGLE-TR-39 & $17\uph57\upm05\zdot\ups70$ & $-29\arcd22\arcm48\zdot\arcs7$ &   0.8152 & 60.81057 & 14.685 & 1.30 & 0.030 &11 &  \\ 
OGLE-TR-40 & $17\uph57\upm10\zdot\ups27$ & $-29\arcd15\arcm38\zdot\arcs1$ &   3.4318 & 60.03089 & 14.947 & 1.25 & 0.026 & 3 &  \\ 
OGLE-TR-41 & $17\uph55\upm16\zdot\ups39$ & $-29\arcd31\arcm32\zdot\arcs1$ &   4.5170 & 62.29677 & 13.488 & 0.65 & 0.022 & 2 & OII \\ 
OGLE-TR-42 & $17\uph55\upm29\zdot\ups76$ & $-29\arcd33\arcm30\zdot\arcs8$ &   4.1610 & 63.67328 & 15.397 & 1.37 & 0.038 & 4 & OII \\ 
OGLE-TR-43 & $17\uph51\upm27\zdot\ups04$ & $-29\arcd52\arcm21\zdot\arcs8$ &    --    &   --     & 15.506 & 0.95 & 0.035 & 1 &  \\ 
OGLE-TR-44 & $17\uph50\upm45\zdot\ups98$ & $-30\arcd03\arcm40\zdot\arcs1$ &    --    &   --     & 14.573 & 1.10 & 0.059 & 1 & OII \\ 
OGLE-TR-45 & $17\uph50\upm48\zdot\ups10$ & $-30\arcd00\arcm38\zdot\arcs7$ &    --    &   --     & 16.073 & 1.28 & 0.062 & 1 & OII \\ 
OGLE-TR-46 & $17\uph54\upm00\zdot\ups88$ & $-29\arcd43\arcm45\zdot\arcs7$ &    --    &   --     & 13.449 & 0.93 & 0.049 & 1 & OII \\ 
\hline}

Light curves in the graphical form are presented in Appendix: full
phased light curve and close-up around the eclipse. Additionally we also
print there a finding chart -- $60\times 60$ arcsec subframe of the {\it
I}-band template image centered on the star. North is up and East to the
left in these images. For single transit stars, only the light curve
around the transit and finding chart are presented.

\Section{Discussion}
Results of our 2001 planetary and low-luminosity object transit campaign
clear\-ly indicate that the OGLE project reached photometric accuracy of
millimagnitudes while observing millions of stars in the most crowded
fields of the Galactic disk and bulge. Transit events with the depth as
small as 10--20 millimagnitude can routinely be detected in the data
collected by the OGLE-III hardware and processed by the OGLE-III
software data pipeline. This opens up a new discovery channel of
low-luminosity companions by photometric monitoring of millions of
stars. We note that with the spectroscopic follow-up observations it
will soon be possible to determine all basic parameters of these faint
objects: planets, brown dwarfs or low-luminosity end of the main
sequence stars.

A limited analysis is possible even now, without spectroscopic data. We
estimate sizes of our transiting objects by modeling their light curves
assuming a completely dark companion and using formulae provided by
Sackett (1999). These were numerically integrated at the appropriate
phase points to produce model light curves. 

We found a possible solution for each candidate with the known period by
analyzing the $\chi^2$ of the model fit \{$R_s/a$, $R_c/R_s$, $i$\} to
the observational data, where $a$ is the semi-major axis of the orbit,
$R_s$ -- radius of the primary star, $R_c$ -- radius of the low
luminosity companion and $i$ -- inclination of the orbit. In all cases
we assumed the limb darkening coefficient $u=0.5$ (Al-Naimiy 1978).

As expected, many fits of similar quality can be obtained for an entire
class of parameters. This corresponds to the fact that very similar
shape of the eclipse can be obtained  by a small companion with orbital
inclination ${i=90\arcd}$ or larger one with smaller $i$ transiting on
the disk of larger star.  However, the ratio of radii remains almost
constant. Thus, if the stellar radius were known, the size of the
companion and the inclination angle $i$ could be derived unambiguously.
The semi-major axis of the orbit can be calculated from the determined
orbital period, assuming that the stellar mass $M_s = 1~\MS$. The colors
of our transit stars indicate that these stars have spectral types
similar to, or later than the Sun.

The smallest size of the star and its companion is obtained when the
transit is central, \ie for the inclination ${i=90\arcd}$. Non-central
passage (smaller $i$) requires a larger size of both, the star and its
companion. Table~3 lists the minimum size of the transiting companion
and the corresponding size of the star to provide a feeling about the 
expected size of the observed objects. It should be remembered that
the semi-major axis of the orbit scales as $ M^{1/3} $, and therefore the
same scaling is appropriate for the sizes of stars and companions listed
in Table~3. In the close-up windows in the Appendix we show model light
curves for central transit (${i=90\arcd}$) drawn with a continuous bold
line.
\renewcommand{\arraystretch}{0.95}
\renewcommand{\TableFont}{\footnotesize}
\MakeTable{lccclcc}{12.5cm}{Dimensions of stars and companions for central
passage ($M_s=1~\MS$)}
{\cline{1-3}\cline{5-7}
\noalign{\vskip3pt}
Name       & $R_s$   & $R_c$ & $\phantom{xxxxxxx}$ & Name & $R_s$ & $R_c$\\
&[\RS]&[\RS]&$\phantom{xxxxx}$ &&[\RS]&[\RS]\\
\noalign{\vskip3pt}
\cline{1-3}\cline{5-7}
\noalign{\vskip3pt}
OGLE-TR-1  &    0.99 &    0.18& & OGLE-TR-22 &    1.68 &    0.42 \\
OGLE-TR-2  &    1.75 &    0.21& & OGLE-TR-23 &    0.99 &    0.21 \\
OGLE-TR-3  &    1.48 &    0.18& & OGLE-TR-24 &    1.45 &    0.29 \\
OGLE-TR-4  &    1.60 &    0.35& & OGLE-TR-25 &    1.43 &    0.24 \\
OGLE-TR-5  &    1.01 &    0.18& & OGLE-TR-26 &    1.15 &    0.23 \\
OGLE-TR-6  &    1.65 &    0.33& & OGLE-TR-27 &    1.88 &    0.26 \\
OGLE-TR-7  &    1.37 &    0.22& & OGLE-TR-28 &    1.76 &    0.35 \\
OGLE-TR-8  &    0.85 &    0.16& & OGLE-TR-29 &    0.87 &    0.17 \\
OGLE-TR-9  &    1.36 &    0.26& & OGLE-TR-30 &    1.38 &    0.23 \\
OGLE-TR-10 &    0.84 &    0.11& & OGLE-TR-31 &    1.60 &    0.24 \\
OGLE-TR-11 &    1.03 &    0.21& & OGLE-TR-32 &    0.88 &    0.14 \\
OGLE-TR-12 &    0.89 &    0.15& & OGLE-TR-33 &    1.26 &    0.20 \\
OGLE-TR-13 &    1.49 &    0.22& & OGLE-TR-34 &    1.92 &    0.36 \\
OGLE-TR-14 &    2.02 &    0.32& & OGLE-TR-35 &    1.12 &    0.17 \\
OGLE-TR-15 &    3.02 &    0.42& & OGLE-TR-36 &    1.46 &    0.31 \\
OGLE-TR-16 &    2.49 &    0.35& & OGLE-TR-37 &    2.24 &    0.34 \\
OGLE-TR-17 &    1.53 &    0.25& & OGLE-TR-38 &    0.95 &    0.18 \\
OGLE-TR-18 &    1.09 &    0.20& & OGLE-TR-39 &    0.97 &    0.14 \\
OGLE-TR-19 &    0.84 &    0.18& & OGLE-TR-40 &    0.73 &    0.10 \\
OGLE-TR-20 &    1.16 &    0.24& & OGLE-TR-41 &    1.55 &    0.20 \\
OGLE-TR-21 &    2.38 &    0.43& & OGLE-TR-42 &    1.55 &    0.26 \\
\cline{1-3}\cline{5-7}}

Table~3 indicates that the companions cover a large range of sizes of
low-luminosity objects. Many of them are certainly faint M-type stars.
In several cases (\eg OGLE-TR-5, OGLE-TR-16), the light curves exhibit
an ellipsoidal variation, indicating that the primary is tidally
distorted, \ie the companion mass is not very small. There are eight
objects (OGLE-TR-8, OGLE-TR-10, OGLE-TR-12, OGLE-TR-29, OGLE-TR-32,
OGLE-TR-35, OGLE-TR-39, OGLE-TR-40) with companion radii of the order of
1.5~${\rm R}_{\rm Jup}$ or less, which implies that they may by planets,
or brown dwarfs or low-luminosity stars.  Note that the radius of the
``hot Jupiter''  orbiting and transiting HD~209458 is about $ 1.4~{\rm
R}_{\rm Jup}$.

The most intriguing and exciting objects within this group are certainly
OGLE-TR-40 and OGLE-TR-10.  Three and four individual transits,
respectively,  were observed in these stars, so the orbital periods are
sound. The radius estimates of the companions as given in Table~3,
$1.0~{\rm R}_{\rm Jup}$ and $1.1~{\rm R}_{\rm Jup}$, respectively, are
the smallest among our systems.  Even if the transits were significantly
non-central the radii of these objects would be smaller than $1.5~{\rm
R}_{\rm Jup}$. These objects appear to be most similar to the ``hot
Jupiter'' companion to HD~209458.

It is worth noting that the objects presented in Table~2 were selected by the 
presence of apparent transits only, with no knowledge on any other their 
properties. Thus, the sample is very well suited for general study of low-
luminosity objects. The spectroscopic follow-up will provide a simple and very 
important clarification.  The radial velocity amplitude should give an 
estimate of the companion mass accurate enough to determine its nature: a 
planet, a brown dwarf, or a low mass star.  The spectrum of the primary will 
provide an estimate of its mass and luminosity, and a refinement of the 
companion mass.  Finally, a very accurate future photometry may refine system 
parameters even more. 

Note, that a detection of a brown dwarf companion on a short period
orbit would be at least as interesting as a detection of Jupiter mass
planet.  There appears to be a ``brown dwarf desert'', with very few
objects in the mass range $10~{\rm M}_{\rm Jup} - 100~{\rm M}_{\rm Jup}$
(\cf Tabachnik and Tremaine 2001, and references therein).  The
spectroscopic follow-up of our 42 objects will either support or
contradict the concept of the ``desert''.

There is a major problem with the mass--radius relation for lower main 
sequence stars, with the observed radii being 10--20\% larger than
model predictions (Torres and Ribas 2001,  O'Brien, Bond and Sion 2001).
There is a plausible explanation in terms of long living huge star spots
(Spruit 1982, Spruit and Weiss 1986), but there are too few accurate
radius determinations for a verification of that theory.  The sample 
of eclipsing systems with lower main sequence secondaries will increase
substantially as soon as the OGLE-III data will be searched for deeper 
eclipses than those presented in this paper.

In addition to 42 stars for which multiple transits were observed, we
found four objects with low luminosity companions, but only one transit
event detected in our dataset.  The lack of additional transits during
our 2001 campaign indicates that the orbital periods of these objects
are likely to be long.

The successful 2001 campaign proves that photometric detection of 
low-lumino\-sity transiting objects is not only feasible but that transits can 
routinely be detected. Therefore it will be worthwhile to repeat similar 
observational campaigns  in other directions of the Galactic disk.  Such 
campaigns are planned in the  coming seasons during the OGLE-III phase of our 
project. 

The photometric data of OGLE-III transit objects  presented in this
paper are available in the electronic form from the OGLE archive: 
\vspace*{-7pt}
\begin{center}
{\it http://www.astrouw.edu.pl/\~{}ogle} \\
{\it ftp://ftp.astrouw.edu.pl/ogle/ogle3/transits}\\
\end{center}
\vspace*{-9pt}
or its US mirror
\vspace*{-11pt}
\begin{center}
{\it http://bulge.princeton.edu/\~{}ogle}\\
{\it ftp://bulge.princeton.edu/ogle/ogle3/transits}\\
\end{center}

\Acknow{The paper was partly supported by the  Polish KBN grant
2P03D01418 to M.\ Kubiak. Partial support to the OGLE project was
provided with the NSF  grant AST-9820314 to B.~Paczy\'nski. We
acknowledge usage of The Digitized Sky Survey which was  produced at the
Space Telescope Science Institute based on photographic data  obtained
using The UK Schmidt Telescope, operated by the Royal Observatory 
Edinburgh.}

\end{document}